\documentclass[final,5p,times,twocolumn,authoryear]{elsarticle}

\usepackage{amssymb, amsthm, amsmath, amsfonts, amsbsy, mathrsfs}
\usepackage{natbib}
\setcitestyle{numbers,square}
\usepackage{hyperref}

\journal{Physics Letters B}

\begin{document}

\begin{frontmatter}

\title{Cosmological constant as an integration constant}

\author[aff1]{Justin C. Feng}
\author[aff1]{Pisin Chen}
\affiliation[aff1]{organization={Leung Center for Cosmology and Particle Astrophysics, National Taiwan University},
            addressline={No.1 Sec.4, Roosevelt Rd.}, 
            city={Taipei},
            postcode={10617}, 
            state={},
            country={Taiwan, R.O.C.}}

\begin{abstract}
The discrepancy between the observed value of the cosmological constant (CC) and its expected value from quantum field theoretical considerations motivates the search for a theory in which the CC is decoupled from the vacuum energy. In this article, we consider the viability of theories in which the Einstein equations are recovered (without additional constraints) and in which the CC is regarded as an integration constant. These theories include trace-free Einstein gravity, theories constructed from the Codazzi equation  (which includes Cotton gravity and a gauge-gravity inspired theory), and conformal Killing gravity. We remark on a recent debate regarding Cotton gravity and find that while the Codazzi equation of that theory is indeed underdetermined, the solutions of the Codazzi equation trivialize to $\lambda g_{ab}$ on generic backgrounds, and that in principle, one can close the system with the divergence-free condition and an appropriate choice of initial data. We also propose a full variational principle (full in the sense that variations in all variables are considered) for each of the aforementioned theories that can incorporate the matter sector; in this manner, we can obtain the trace-free Einstein equations without a unimodular constraint. The resulting actions require additional (auxiliary) fields and are therefore only expected to be effective, but they may provide a useful starting point in bottom up approaches to constructing more fundamental theories.
\end{abstract}


\end{frontmatter}

\section{Introduction}
\label{introduction}

Since Einstein introduced the cosmological constant (CC) to his equations in 1917 \cite{Einstein:1917}, the debate about its nature has never stopped. Before the discovery of the accelerating expansion of the universe in 1999 \cite{SupernovaSearchTeam:1998fmf,SupernovaCosmologyProject:1998vns}, the main consideration was whether the CC is exactly zero (See, for example, the famous review by Weinberg \cite{Weinberg:1988cp}). In the aftermath of the discovery of the accelerating expansion of the universe, the problem became more acute. The notion of dark energy introduced to account for this effect behaves like the CC. Indeed, in most analyses in cosmology, such as that of the cosmic microwave background, the CC is commonly invoked in the modeling. If true, then the observed value for the CC, fixed by observations of the accelerating expansion of the universe \cite{Chen:2010at}, is many orders (up to 120) of magnitude smaller than contributions from vacuum fluctuations arising from quantum field theoretic considerations. The discrepancy between the vacuum energy and the observed value for the CC requires an extreme degree of fine tuning. On the other hand, the Einstein field equations have been well-established on astrophysical scales, even in the strong gravity regime. This state of affairs strongly motivates a modification or reconstruction of Einstein gravity.

In this article, we consider one strategy, discussed in Section VII of \cite{Weinberg:1988cp}, in which one seeks a formulation of Einstein gravity in which the CC arises as a constant of integration in a manner independent of vacuum fluctuations, avoiding the need for fine tuning in this regard. Though the most well-known of these approaches is that of unimodular gravity \cite{Anderson:1971pn,vanderBij:1981ym,Buchmuller:1988wx,Buchmuller:1988yn,Henneaux:1989zc,Weinberg:1988cp} (see \cite{Carballo-Rubio:2022ofy,Alvarez:2023utn} for recent overviews of the topic), which can be formulated at the level of the action, unimodular gravity introduces a constraint on the metric determinant that breaks general covariance. On the other hand, at the level of the field equations, there are several proposals in the literature that do not require such a constraint; our main aim in this article will be to evaluate the viability of these approaches. 

A proposal originating from Einstein \cite{Einstein:1919} is that of the trace-free Einstein equations, which are recovered in unimodular gravity, but can be considered on their own without the unimodular constraint \cite{Ellis:2010uc,deCesare:2021wmk}. Another proposal by Cook and Chen \cite{Cook:2008mx,Chen:2010at}, inspired by a set of third-order equations developed by Kilmister, Newman \cite{Kilmister:1959gqq,Kilmister:1961}, and later by Yang \cite{Yang:1974kj} is to construct a gravity theory from the Bianchi identities and derivatives of the energy-momentum tensor. We focus primarily on the aforementioned approaches, though we mention here other proposals in which one may obtain the unconstrained Einstein field equations with the CC arising as an integration constant, such as the thermodynamical approach of Padmanabhan \cite{Padmanabhan:2014nca,Padmanabhan:2016eld} (see also \cite{Dadhich:2016vbn}), a modified metric-affine theory \cite{Tapia:1996ue,Tapia:1998qt,Tapia:1998as}, and contributions from a three-form action \footnote{However, we note that in this case the vacuum energy still contributes to the cosmological term.} \cite{Duff:1980qv,Henneaux:1984ji}. 

More recently, Harada proposed two theories of gravity, termed Cotton gravity \cite{Harada:2021bte} and the so-called conformal Killing gravity \cite{Harada:2023rqw,Mantica:2023stl}. Cotton gravity is constructed from the Cotton tensor, and it has been shown that one can rewrite the field equations of Cotton gravity as a Codazzi equation $\nabla_{[a}C_{b]c}=0$ \cite{Mantica:2022flg,Sussman:2023eep}; in this regard, Cotton gravity is in the same class as the earlier theory of Cook and Chen, as we will show that the field equations for the latter can also be reformulated as a Codazzi equation. Conformal Killing gravity, has a different structure altogether, as the field equations resemble the defining equation for a conformal Killing tensor.

Cotton gravity was recently the focus of a debate \cite{Clement:2023tyx,Sussman:2024iwk,Clement:2024pjl,Sussman:2024qsg} over its viability as a physical theory, as it was discovered that the symmetry-reduced equations of Cotton gravity are underdetermined. It is been argued in \cite{Sussman:2024iwk,Sussman:2024qsg} that while the equations of Cotton gravity are indeed underdetermined in cases of high symmetry (for instance, cosmological or spherically symmetric spacetimes), this is not necessarily the case in generic spacetimes. Similar concerns apply to the theory of Cook and Chen, as they share the same Codazzi structure. Another concern that applies to Codazzi type theories and conformal Killing gravity, is the fact that they are only defined at the level of the field equations; these theories suffer from the lack of a complete variational principle, as the variational principles proposed so far require holding some of the variables (such as the spacetime metric $g_{ab}$) fixed. One difficulty is that the energy-momentum tensor is obtained from variations of the metric tensor in the matter sector, but derivatives of the energy-momentum tensor appear in the field equations. 

As remarked earlier, the primary aim of this article is to assess the viability of the aforementioned approaches to decoupling the CC from the vacuum energy (without introducing additional constraints). We focus on two specific requirements. First, we require that the field equations can be reformulated in a manner that yields a well-posed initial value problem---in doing so, we aim to shed light on the recent debate over the Codazzi equation. Second, we propose variational principles (including matter) for the trace-free Einstein equations, the Codazzi equation, and the conformal Killing equation without introducing additional constraints. It should be noted that some variational principles for the trace-free Einstein equations were recently proposed \cite{Montesinos:2023pjp,Montesinos:2024yjs} that avoid additional constraints, but our approach is the only one so far to explicitly incorporate the matter sector while avoiding the need for fine-tuning. Though our proposed actions are not intended to be fundamental, as they require the introduction of auxiliary fields and ghosts (which do not exhibit classical pathologies), they may serve as a first step toward a more complete theory. 

\section{Theories}


The theories we consider here differ significantly in their construction, but they all share the property that they yield the unconstrained Einstein equations in a manner that is entirely independent of the vacuum energy contribution in the energy-momentum tensor (avoiding the need for fine-tuning), with the cosmological constant term arising as a constant of integration. In this section, we briefly review the aforementioned theories.

\subsection{Trace-free Einstein equations}

The trace-free Einstein equations may be obtained by assuming that the determinant of the metric is held fixed under the variation of the action; in this form, the theory is called unimodular gravity. However, the equations themselves can be written without assuming such a constraint \cite{Ellis:2010uc,deCesare:2021wmk}:
\begin{equation} \label{Eq:TFEFE}
    R_{\mu\nu}-\frac{1}{4}R g_{\mu\nu} = \kappa (T_{\mu\nu}-\frac{1}{4}T g_{\mu\nu}) , \qquad \nabla^\mu T_{\mu\nu}=0,
\end{equation}
where $\underline{R}_{\mu\nu}$ is the Ricci tensor and $\nabla^\mu T_{\mu\nu}=0$ is imposed as an independent condition on the energy-momentum tensor. The contracted Bianchi identity $\nabla^\mu (R_{\mu\nu}-\tfrac{1}{2}R g_{\mu\nu})=0$ combined with Eq.\eqref{Eq:TFEFE} implies $\nabla_\mu(R+\kappa T)=0$, which has the solution $R+\kappa T=4\bar\Lambda$, where $\bar\Lambda$ is an integration constant. Solving for $T$ and plugging this back into Eq. \eqref{Eq:TFEFE} yields:
\begin{equation} \label{Eq:EFEcc}
    R_{\mu\nu}-\frac{1}{2}R g_{\mu\nu} + \bar\Lambda g_{\mu\nu} = \kappa T_{\mu\nu}.
\end{equation}
Later, we consider a variational principle that does not impose a unimodular constraint on the metric.


\subsection{Bianchi gravity}

The works \cite{Cook:2008mx,Chen:2010at} consider a field theory in
which the connection and curvature tensors are treated as analogues of
the respective gauge field and field strength tensor of
electrodynamics, yielding:
\begin{equation} \label{Eq:Bianchi}
\nabla_{\rho} R_{\mu\nu\sigma\tau}
+
\nabla_{\tau} R_{\mu\nu\rho\sigma}
+
\nabla_{\sigma} R_{\mu\nu\tau\rho}
=0
\end{equation}
\begin{equation} \label{Eq:FieldEquation}
    \nabla_{\sigma} R^{\sigma\rho}{}_{\mu\nu}
    =\tfrac{\kappa}{2} J_{\mu\nu}{}^{\rho}
\end{equation}

\noindent where Eq. \eqref{Eq:Bianchi} is the Bianchi identity and $R_{\mu\nu\sigma\tau}$ is the curvature tensor. One postulates the source $J_{\mu\nu}{}^{\rho}$ to be
integrable, so that for some symmetric rank-2 tensor $\bar{T}_{\mu\nu}$,
it may be written as:
\begin{equation} \label{Eq:IntegrableJ}
    J_{\mu\nu}{}^{\rho} = -q \nabla_{[\mu} \bar{T}_{\nu]}{^\rho}.
\end{equation}
The field equation \eqref{Eq:FieldEquation} may be rewritten:
\begin{equation} \label{Eq:FieldEquation2}
    {\nabla}_{[\mu}R_{\nu]}{}{^\rho} - {\kappa} \nabla_{[\mu} \bar{T}_{\nu]}{^\rho} =0,
\end{equation}
where $\bar{T}_{\mu\nu}={T}_{\mu\nu}-\tfrac{1}{2}g_{\mu \nu} T$ is identified with the trace-reversal of the energy-momentum tensor. One class of solutions for Eq. \eqref{Eq:FieldEquation2} corresponds
to the trace-reversed Einstein equation, up to some integration tensor
$X_{\mu\nu}$:
\begin{equation} \label{Eq:FieldEquationTrRev}
    R_{\mu\nu}
    =\kappa ({T}_{\mu\nu}-\tfrac{1}{2}g_{\mu \nu} T+{X}_{\mu\nu}),
\end{equation}
where $X_{\mu\nu}$ satisfies the following:
\begin{equation} \label{Eq:intTensor}
    X_{\mu\nu}=X_{\nu\mu}, \qquad \nabla^\mu X_{\mu\nu} =0
    , \qquad \nabla_{[\mu} X_{\nu]\rho} =0.
\end{equation}
The tensor $X_{\mu\nu}$ is assumed to be determined by
boundary and initial data; if the initial/boundary data yields $X_{\mu\nu}=\lambda
g_{\mu\nu}$, then the CC $\lambda$ may be interpreted
as an integration constant (we discuss later how such a solution arises from initial data).

One might imagine rewriting Eq. \eqref{Eq:FieldEquation2} as:
\begin{equation} \label{Eq:FieldEquationCodazziForm}
    {\nabla}_{[\mu}K_{\nu]}{}{^\rho}=0,
\end{equation}
where:
\begin{equation} \label{Eq:FieldEquationTrRevKTens}
    K_{\mu\nu}
    =R_{\mu\nu} - \kappa ({T}_{\mu\nu}-\tfrac{1}{2}g_{\mu \nu} T+{X}_{\mu\nu}).
\end{equation}
Equation \eqref{Eq:FieldEquationCodazziForm} has the form of a Codazzi equation \cite{Mantica:2022flg,Sussman:2023eep}, which is of recent interest in the Cotton gravity program \cite{Harada:2021bte}.


\subsection{Cotton gravity}

Cotton gravity is based on the Cotton tensor:
\begin{equation} \label{Eq:CottonTensor}
    C_{abc}
    :=\nabla_b R_{ac} - \nabla_c R_{ab} - \tfrac{1}{6} \left(g_{ac}\nabla_b R - g_{ab} \nabla_c R\right)
\end{equation}
Defining a new tensor $\mathscr{T}_{abc}$
\begin{equation} \label{Eq:EMTensor}
    \mathscr{T}_{abc}
    :=\nabla_b T_{ac} - \nabla_c T_{ab} - \tfrac{1}{6} \left(g_{ac}\nabla_b T - g_{ab} \nabla_c T\right),
\end{equation}
one can write the field equation for Cotton gravity as:
\begin{equation} \label{Eq:CottonGravityEquation}
    C_{abc} = 2\kappa \mathscr{T}_{abc}.
\end{equation}
This equation can be rewritten in the form:
\begin{equation} \label{Eq:CodazziFormCottonGravity}
    {\nabla}_{[a}\mathcal{C}_{b]}{}{^c}=0,
\end{equation}
where (with $X_{ab}$ satisfying Eq. \eqref{Eq:intTensor}):
\begin{equation} \label{Eq:CottonGravityCTensor}
    \mathcal{C}_{ab} = G_{ab} - \kappa T_{ab} - \frac{1}{3}(G- \kappa T) g_{ab} + X_{ab}.
\end{equation}


\subsection{`Conformal'' Killing gravity}

The equations for the so-called conformal Killing gravity theory proposed in \cite{Harada:2023rqw} may be written in the form \cite{Mantica:2023stl}:
\begin{equation} \label{Eq:ConformalKillinggravity}
\begin{aligned}
    &6 (K_{bc;a} + K_{ca;b} + K_{ab;c}) =  g_{bc} K_{,a} + g_{ca} K_{,b} + g_{ab} K_{,c},
    \\
    &\qquad\qquad\qquad\qquad\qquad\qquad\qquad \\
    &K_{ab}:=\kappa T_{ab} - G_{ab}.
\end{aligned}
\end{equation}
The equation above has the form of the equation for a conformal Killing tensor:
\begin{equation} \label{Eq:ConformalKillingTensor}
    K_{bc;a} + K_{ca;b} + K_{ab;c} = u_{a} g_{bc} + u_{b} g_{ca} + u_{c} g_{ab},
\end{equation}
where $u_a$ is a one-form; $K_{ab}$ is a symmetric Killing tensor when $u_a=0$. However, as pointed out in \cite{Barnes:2024gko,Barnes:2023uru} a simple redefinition $\bar{K}_{ab}:=6K_{ab}-g_{ab} K$ reveals that any nontrivial solution of Eq. \eqref{Eq:ConformalKillinggravity} requires the existence of a Killing tensor given by $\bar{K}_{ab}$. It follows that the solutions of Eq. \eqref{Eq:ConformalKillinggravity} either trivialize to $K_{ab}=-\lambda g_{ab}$, or admit a Killing tensor. For this reason, it is perhaps more appropriate to refer to this theory as ``Killing tensor gravity'', but we retain here the common terminology in the literature. Moreover, we note that perturbations of Killing tensor solutions satisfying $K_{ab}\neq-\lambda g_{ab}$ must also admit a Killing tensor; since our universe (being inhomogeneous and anisotropic on small scales) does not have the required symmetries to admit a global Killing tensor, we regard such solutions as unphysical.


\section{Codazzi equation}

Equations \eqref{Eq:FieldEquationCodazziForm} and \eqref{Eq:CodazziFormCottonGravity} have the form of a Codazzi equation:
\begin{equation} \label{Eq:CodazziFieldEquation}
    {\nabla}_{[a}C_{b]}{}{^c}=0,
\end{equation}
where $C_{ab}=C_{ba}$ is termed a Codazzi tensor, following the terminology of \cite{Mantica:2022flg,Sussman:2023eep}. One can perhaps imagine a class of theories defined by an equation of the form in Eq. \eqref{Eq:CodazziFieldEquation}. Now there is some debate over whether equations of this type are predictive \cite{Clement:2023tyx,Sussman:2024iwk,Clement:2024pjl,Sussman:2024qsg}; in particular, it has been claimed in \cite{Clement:2024pjl} that the Codazzi equations are underdetermined, and in a responding comment it was claimed that this only occurs for highly symmetric spacetimes \cite{Sussman:2024qsg}. In the following, we attempt to shed some light on this dispute.

Contracting Eq. \eqref{Eq:CodazziFieldEquation}, one obtains:
\begin{equation} \label{Eq:CodazziFieldEquationContract}
    {\nabla}_{c}C_{a}{}{^c}-{\nabla}_{a}C=0.
\end{equation}
where $C:=C_{a}{}{^a}$. Taking the divergence of Eq. \eqref{Eq:CodazziFieldEquation}, one obtains a wave-like equation:
\begin{equation} \label{Eq:CodazziFieldEquationDivA}
    \Box C_{a}{}{^b}-\nabla_a{\nabla}_{c}C{}^{c b}-R_{ac}C^{cb}-R_{acd}{^b} C^{cd}=0.
\end{equation}
and the divergence of Eq. \eqref{Eq:CodazziFieldEquationContract} is:
\begin{equation} \label{Eq:CodazziFieldEquationContractDiv}
    \Box C - {\nabla}_{a}{\nabla}_{b}C^{ab}=0.
\end{equation}
Of course, Eq. \eqref{Eq:CodazziFieldEquationDivA} is underdetermined due to the term $\nabla_a{\nabla}_{c}C{}^{c b}$. In an orthonormal basis, the derivative $\partial^2_t C{_t}^b$ disappears from Eq. \eqref{Eq:CodazziFieldEquationDivA}, so that the time evolution of the components $C{_t}^b$ is ill-defined---the equations are indeed underdetermined. However, Eq. \eqref{Eq:CodazziFieldEquationDivA} reveals a straightforward resolution; one should supply the Codazzi equation with a constraint equation, and a natural one is the divergence-free condition $\nabla_a C^{ab}=0$, which is essentially a statement that $C^{ab}$ satisfies a local conservation law.\footnote{This is natural in the sense that each term in Eq. \eqref{Eq:CottonGravityCTensor} for $C_{ab}$ independently satisfies a conservation law when the Einstein equation holds (up to a cosmological term, also cf. Eq. \eqref{Eq:intTensor}) and in particular when the energy-momentum tensor satisfies the local conservation law $\nabla_a T^{ab}=0$.} Of course, an attempt to directly solve the Codazzi equations supplemented with the constraint $\nabla_a C^{ab}=0$ may still lead to ambiguities, but such ambiguities correspond to a partial freedom (which we restrict in the next paragraph) in choosing initial data for $C_{ab}$ and $\partial_t C_{ab}$ for the system:
\begin{equation} \label{Eq:CodazziFieldEquationDivAeqsys}
\begin{aligned}
    \Box C_{a}{}{^b}-R_{ac}C^{cb}-R_{acd}{^b} C^{cd}=0,
    \qquad
    \nabla_a C^{ab}=0.
\end{aligned}
\end{equation}

While solutions of Eq. \eqref{Eq:CodazziFieldEquation} satisfying the divergence-free constraint will also satisfy the system in Eq. \eqref{Eq:CodazziFieldEquationDivAeqsys}, there is no guarantee that solutions of Eq. \eqref{Eq:CodazziFieldEquationDivAeqsys} will satisfy Eq. \eqref{Eq:CodazziFieldEquation}. To identify the conditions on the initial data that allow one to recover solutions of the Codazzi equation, we consider the application of the D'Alembertian to the following quantity:
\begin{equation} \label{Eq:QDefn}
    Q_{ab}{^c}:=2{\nabla}_{[a}C_{b]}{}{^c},
\end{equation}
which after using Eq. \eqref{Eq:CodazziFieldEquationDivA} has the form:
\begin{equation} \label{Eq:CodazziProp}
    \begin{aligned}
        \Box Q_{cde}
        =\,&
        R_{cdeb} \nabla _{a}C^{ab} + 2(\nabla ^{b}C^{a}{}_{[c})R_{d]bea} + 4 \nabla _{[e}R_{a][c} C_{d]}{}^{a}\\
        &
        + 2 C^{ab} \nabla _{[d}R_{c]aeb}- Q_{d}{}^{a}{}_{e} R_{ca} - Q_{ace} R_{d}{}^{a}\\
        &
         - 2 Q^{ab}{}_{e} R_{cadb} + Q_{d}{}^{ab} R_{caeb} - Q_{c}{}^{ab} R_{daeb}.
    \end{aligned}
\end{equation}
Note that the right-hand side contains at most first derivatives of $C_{ab}$ (and also that the divergence-free condition removes the first term). One can then regard Eq. \eqref{Eq:CodazziProp} as a wave equation for $Q_{cde}$, with $C_{ab}$ being an independent variable satisfying Eq. \eqref{Eq:CodazziFieldEquationDivA} and the constraint \eqref{Eq:CodazziProp}. Now consider initial data consistent with Eq. \eqref{Eq:CodazziFieldEquation}, so that $Q_{cde}=0$ (but not its derivatives). For nonvanishing curvature, the right-hand side will generically produce a source unless $C_{ab}=\lambda g_{ab}$ on a Cauchy hypersurface and $\nabla_c C_{ab}=0$. This indicates that if $C_{ab} \neq \lambda g_{ab}$ and $\nabla_c C_{ab} \neq 0$ satisfies Eq. \eqref{Eq:CodazziProp} on a Cauchy surface, the time evolution of $C_{ab}$ under Eq. \eqref{Eq:CodazziFieldEquationDivA} will in general generate violations of Eq. \eqref{Eq:CodazziProp}. There may also be special cases (highly symmetric spacetimes, for instance) in which Eq. \eqref{Eq:CodazziFieldEquation} may admit nontrivial solutions $C_{ab} \neq \lambda g_{ab}$, but one cannot expect the existence of solutions to Eq. \eqref{Eq:CodazziFieldEquation} for generic initial data. Observe that Eq. \eqref{Eq:CodazziProp} and the conclusions outlined in the previous paragraph are independent of the constraint $\nabla_a C^{ab}=0$. 

To summarize, the Codazzi equation, supplied with the divergence-free constraint $\nabla_a C^{ab}=0$, implies that $C_{ab}$ satisfies Eq. \eqref{Eq:CodazziFieldEquationDivAeqsys}, which has a well-defined initial value problem. While Eq. \eqref{Eq:CodazziFieldEquationDivAeqsys} also admits solutions that do not satisfy the Codazzi equation, one can restrict the space of initial data to eliminate such solutions that remain. The preceding analysis surrounding \eqref{Eq:CodazziProp} indicates that in a general spacetime, one can only guarantee the existence of solutions to the Codazzi equation for initial data also satisfying $C_{ab}=\lambda g_{ab}$ and $\nabla_c C_{ab}=0$, where $\lambda$ is an arbitrary constant; other solutions will likely require background geometries with special properties, as argued in \cite{Sussman:2024iwk,Sussman:2024qsg}. We find that both sets of authors, \cite{Clement:2023tyx,Clement:2024pjl} and \cite{Sussman:2024iwk,Sussman:2024qsg}, present valid points: theories based on the Codazzi tensors are indeed underdetermined, but the space of solutions is nonetheless highly restricted in generic settings.


\section{Variational principle}

The variational principles presented in the literature for the trace-free Einstein equations \cite{Weinberg:1988cp} and theories admitting a Codazzi equation formulation (Bianchi gravity \cite{Cook:2008mx,Chen:2010at} and Cotton gravity \cite{Harada:2021bte}) typically require holding certain quantities fixed; for the trace-free Einstein equations, the metric determinant is held fixed (corresponding to an additional constraint), and for the Codazzi equation, the metric is held fixed (as variations in the metric will generally introduce additional unwanted constraints). We therefore seek a general variational principle that yields the desired field equations (including contributions from the matter sector) without unwanted constraints under a full variation---that is, a variation with respect to all variables appearing in the action.

Consider a simple action of the form for the case of the Codazzi equation:
\begin{equation} \label{Eq:ActionSC}
    \begin{aligned}
    S_C:=&\int_\Omega \sqrt{|g|}d^4x [Y^{abc} \nabla_{[a} C_{b]c} + \Psi_{abcdef} Y^{abc} Y^{def}].
    \end{aligned}
\end{equation}
where $Y{^{ab}}_c$ is an auxiliary field, and $\Psi_{abcdef}=\Psi_{defabc}$ is a Lagrange multiplier\footnote{$\Psi_{abcdef}$ may be regarded as a metric on the field space for $Y^{abc}$.}, and $C_{ab}:=G_{ab}-\kappa Z_{ab}$, where $Z_{ab}$ is a field that we require to be equal on shell to the energy-momentum tensor obtained from some matter action $S_M$. The variation of $\Psi_{abcdef}$ yields $Y^{abc}=0$, and the variation with respect to 
$Y^{abc}$ yields:
\begin{equation} \label{Eq:SCCodazzi}
    \nabla_{[a} C_{b]c} + 2\Psi_{abcdef} Y^{def} = 0,
\end{equation}
which reduces to the Codazzi equation when $Y^{abc}=0$. The variation of the action $S_C$ with respect to $g^{ab}$ and $Z_{ab}$ is rather nontrivial, but ultimately consists of terms proportional to products of $Y^{abc}$ and its derivatives; variations with respect to $g^{ab}$ and $Z_{ab}$ vanish when $Y^{abc}=0$. To independently enforce the divergence-free constraint, one may add to the action the following:
\begin{equation} \label{Eq:ActionSD}
    \begin{aligned}
    S_D:=&\int_\Omega \sqrt{|g|}d^4x [U^{b} \nabla^{a} C_{a b} + \psi_{ab} U^{a} U^{b}],
    \end{aligned}
\end{equation}
where $U^a$ is an auxiliary field and $\psi_{ab}=\psi_{ba}$ is a Lagrange multiplier. The variation of $S_D$ with respect to $\psi_{ab}$ implies $U^a=0$, and the variation with respect to $U^a$ yields:
\begin{equation} \label{Eq:SDDF}
    \nabla^{a} C_{a b} + 2\psi_{ab} U^a = 0.
\end{equation}
When $U^a=0$, the above reduces to the divergence-free condition, and the variation of $S_D$ with respect to the remaining variables vanishes. 

It is straightforward to apply this strategy to the other theories considered in this article. For conformal Killing gravity, we construct the action:
\begin{equation} \label{Eq:ActionSC2}
    \begin{aligned}
    S_K:=&\int_\Omega \sqrt{|g|}d^4x [Y^{abc} \nabla_{(a} \bar{K}_{bc)} + \Psi_{abcdef} Y^{abc} Y^{def}].
    \end{aligned}
\end{equation}
where $\bar{K}_{ab}:=6K_{ab}-g_{ab} K$. For the trace-free Einstein equations, we have:
\begin{equation} \label{Eq:ActionTFEFE}
    \begin{aligned}
    S_{TF}:=&\int_\Omega \sqrt{|g|}d^4x [Y^{ab} \left(
        \underline{R}_{ab} - \kappa \underline{Z}_{ab}
    \right) + \Psi_{abcd} Y^{ab} Y^{cd}],
    \end{aligned}
\end{equation}
with $\underline{R}_{ab}:=R_{ab}-\tfrac{1}{4}R g_{ab}$ and $\underline{Z}_{ab}:=R_{ab}-\tfrac{1}{4}Z g_{ab}$. We emphasize that no constraint on the metric determinant is required.

The matter sector is a somewhat delicate matter. Naively, one can add a term to the matter action:
\begin{equation} \label{Eq:AugmentedMatterAction}
    S'_M=S_M+\int_\Omega \sqrt{|g|}d^4x (\tilde{Z}_{ab} g^{ab}),
\end{equation}
so that the variation with respect to $g^{ab}$ yields $\tilde{Z}_{ab}=T_{ab}-\tfrac{1}{2} T g_{ab}$. However, the difficulty with such a strategy is that the variation $\delta \tilde{Z}_{ab}$ will place a nontrivial constraint on $g^{ab}$ unless it becomes a surface term $\delta \tilde{Z}_{ab}=\nabla_c \phi^{c}_{ab}$, or otherwise trivializes. However, the only known tensor with this property is a Ricci tensor formed from a metric-compatible connection, in which case we recover (by way of a Palatini variation) the Einstein equations with a specified CC, or the corresponding Einstein-Cartan equations if fermions are included.

Consider the action:
\begin{equation} \label{Eq:MatterActionmu}
    S_{\bar{M}}=\int_\Omega \sqrt{|g|}d^4x \biggl[  (g^{ab}-h^{ab}) Z_{ab} + (L_h-L_g) \mathcal{E} + \Xi + J\cdot\zeta\cdot J \biggr],
\end{equation}
where $\phi$ is a matter field (assumed to be nonspinorial and minimally coupled), $\chi$ is an extra field that is equal to $\phi$ on shell; we note that $\chi$ is necessarily a ghost, but we argue later that classically, it does not exhibit runaway behavior. Now $h^{ab}$ and $J$ are auxiliary fields, $\zeta$ is a Lagrange multiplier\footnote{In particular, $\zeta$ is defined so that $J \cdot\zeta\cdot J$ is a scalar.}, and the following have been defined:
\begin{equation} \label{Eq:Lags}
    \begin{aligned}
    L_h:=L(\chi,\partial_a\chi,h^{ab}),
    \qquad
    L_g:=L(\varphi,\partial_a\varphi,g^{ab}),\qquad\quad
    \\
    \Xi :=(\varphi-\chi)\cdot J,\qquad
    \mathcal{E}(g^{ab},\underline{h}_{ab})
    :=
    \exp\left[\sqrt{g^{ab}\underline{h}_{ab}}-2\right].
    \end{aligned}
\end{equation}
The variation with respect to $Z_{ab}$ yields $h^{ab}=g^{ab}$, the variation with respect to $\zeta$ yields $J=0$, and the variation with respect to $J$ yields $\varphi=\chi$ when evaluated on $J=0$. The variation of $\mathcal{E}$ has the form:
\begin{equation} \label{Eq:EFfuncVar}
    \delta\mathcal{E} = \frac{\exp\left\{\sqrt{g^{ab}\underline{h}_{ab}}-2\right\}}{\sqrt{g^{ab}\underline{h}_{ab}}}\biggl[\delta g^{ab}\underline{h}_{ab} -g^{cd}\underline{h}_{ac}\underline{h}_{bd} \delta h^{ab}\biggr],
\end{equation}
which on $h^{ab}=g^{ab}$ simplifies to:
\begin{equation} \label{Eq:EFfuncVars}
    \delta\mathcal{E}
    =\frac{1}{2}g_{ab} \left(\delta g^{ab} - \delta h^{ab}\right),
\end{equation}

\noindent Similarly, the Lagrangian variations satisfy:
\begin{equation} \label{Eq:LagsVar}
    \begin{aligned}
    \delta L_h &= \frac{\partial L_h}{\partial h^{ab}} \delta h^{ab} + \frac{\partial L_h}{\partial \chi} \delta \chi + \frac{\partial L_h}{\partial (\partial_a\chi)} \delta \partial_a\chi,\\
    \delta L_g &= \frac{\partial L_g}{\partial g^{ab}} \delta g^{ab} +\frac{\partial L_g}{\partial \varphi} \delta \varphi +\frac{\partial L_g}{\partial (\partial_a\varphi)} \delta \partial_a\varphi.
    \end{aligned}
\end{equation}
One finds that the variations with respect to $g^{ab}$ and $h^{ab}$ yield two constraints that coincide when evaluating on $g^{ab}=h^{ab}$ and $\varphi=\chi$:
\begin{equation} \label{Eq:hgvar}
    \begin{aligned}
    Z_{ab}=\frac{\partial L_h}{\partial h^{ab}} - \frac{1}{2} h_{ab} L_h=\frac{\partial L_g}{\partial g^{ab}} - \frac{1}{2} g_{ab} L_g,
    \end{aligned}
\end{equation}
which makes use of the fact that for $g^{ab}=h^{ab}$ and $\varphi=\chi$, the integrand of Eq. \eqref{Eq:MatterActionmu} is constructed to vanish. The variations with respect to $\varphi$ and $\chi$ yield matter field equations that coincide on $g^{ab}=h^{ab}$ and $\varphi=\chi$, and reduce to the usual field equations obtained from $L_g$ after applying the constraint $J=0$.


\section{Discussion}

In this article, we considered some theories in which the CC arises as an integration constant and yields the Einstein equation without additional constraints. This class includes trace-free Einstein gravity, the Codazzi equation, and conformal Killing gravity. An interesting question, which we have not considered, is whether there are other theories that satisfy this property; we have made an attempt to consider all the proposals in the literature, but the existence and construction of other theories sharing this property (which may perhaps avoid the properties discussed in the following paragraphs) is an interesting question for future investigation.

We revisited a recent debate \cite{Clement:2023tyx,Sussman:2024iwk,Clement:2024pjl,Sussman:2024qsg} concerning the issue of nonuniqueness of solutions to equations of Cotton gravity in symmetry-reduced settiongs. By taking additional derivatives of the Codazzi equations, we find that solutions of the Codazzi equations must satisfy at least one of two conditions: either the solutions coincide with that of the Einstein equations (with arbitrary CC), or that certain contractions of the curvature tensor, the Codazzi tensor, and their derivatives must vanish (in particular, all terms on the right-hand side of Eq. \eqref{Eq:CodazziProp} must vanish). The latter can be satisfied in situations of high symmetry, even if the solutions do not satisfy the Einstein equations, but if we regard such non-Einstein solutions as physical, the right-hand side of Eq. \eqref{Eq:CodazziProp} must vanish exactly even under perturbations. It is unlikely that such conditions can be satisfied unless the perturbations are highly restricted, and one may in this manner argue that such non-Einstein solutions are unphysical. We find that the Codazzi equations can in principle be closed with the addition of the divergence-free constraint so that such non-Einstein solutions can be excluded at the level of initial data.

We formulated a rather general variational principle that can yield the trace-free Einstein equations, the Codazzi equation, and the equations of conformal Killing gravity. An action for trace-free gravity without a unimodular constraint \cite{Montesinos:2023pjp} was recently proposed (another possible approach is described in \cite{Tapia:1996ue,Tapia:1998qt,Tapia:1998as}), but the matter sector was not explicitly incorporated in their treatment; our action \eqref{Eq:ActionTFEFE} provides an alternative that can readily incorporate the matter sector via the action \eqref{Eq:MatterActionmu}. The appearance of Lagrange multipliers and auxiliary fields in the actions amounts to introducing additional nondynamical degrees of freedom---one might prefer an action that depends exclusively on dynamical degrees of freedom. Moreover, there is some question of whether auxiliary fields in gravity theories are pathological \cite{Pani:2012qd,Pani:2013qfa,Ventagli:2023moe}. Regarding the latter, there may be instances in which some of the pathologies (particularly those associated with sharp density gradients) can be mitigated \cite{Kim:2013nna,BeltranJimenez:2017doy,Feng:2022rga}, though it is well-known that one can express the Lagrangian for $f(R)$ theories in the O'Hanlon \cite{OHanlon:1972xqa} form $\psi R-V(\psi)$ (see also 
\cite{Rodrigues:2011zi,DeFelice:2010aj,Sotiriou:2008ve,Nojiri:2010wj,Nojiri:2017ncd} 
and references therein), where $\psi$ is an auxiliary field, and one can easily recover the $f(R)$ Lagrangian from the O'Hanlon Lagrangian by algebraic elimination. Of course, a naive algebraic elimination would trivialize the actions considered here, but the point is that auxiliary fields can provide a starting point for constructing an action exclusively from dynamical degrees of freedom. 

From a more fundamental perspective, one regard auxiliary fields and Lagrange multipliers as vestiges of heavy dynamical degrees of freedom in a more fundamental theory. From this perspective, the actions we propose are not fundamental, but are effective actions obtained after integrating out heavy  dynamical degrees of freedom (see, for instance \cite{deAlwis:2005tg}). In this manner, we are in principle trading an finely tuned cancellation for a scenario with additional heavy degrees of freedom---this is not an unreasonable stance, since from quantum gravity considerations, one expects new physics at scales approaching the Planck scale. In this manner, one may view our variational principle as a bottom-up\footnote{An interesting question for future investigation is whether such a variational strategy is compatible with the complementary top-down approach of the Swampland program \cite{vanBeest:2021lhn,Palti:2019pca}.} strategy for constructing theories in which certain parameters (the CC in our case) are entirely insensitive to short-distance physics.
While the variational principle we have proposed is not fundamental---indeed, one can construct a variational principle for virtually any\footnote{Presumably one could attempt to use the same strategy to construct a variational principle for inconsistent field equations, but this will result in an action that does not admit extrema.} set of consistent field equations in this manner---it may provide a useful first step towards the construction of a more fundamental action for these theories. 

Ghosts are a concern; for instance, the Codazzi equation contains up to three derivatives of the metric. One might also recognize that the duplicate matter degrees of freedom $\chi$ in Eq. \eqref{Eq:MatterActionmu} are necessarily ghosts. At the classical level, there is no issue in the matter sector, as the matter field equations ultimately reduce to those identical to that obtained in the standard variational principle, and the ghost fields are constrained to be equal to the original matter fields---there is no runaway behavior (see also \cite{Deffayet:2021nnt} for another way to avoid runaway behavior). In the gravitational sector, one can place restrictions on the initial data such that the system becomes dynamically equivalent to the Einstein equations with a CC. Regarding the Hamiltonian, we note that on shell, both the matter and gravitational actions vanish in our approach, and it is not too difficult to convince oneself that the total matter Hamiltonian must also vanish on shell. Though there is no issue at the classical level, the appearance of ghosts may be of particular concern for the quantization of the theory. Quantization is beyond the scope of the present article (see the discussion in \cite{Chen:2010at} regarding ghosts in higher curvature theories), but we offer a couple of remarks. First, as argued in the preceding paragraph, the action presented in this article may require further modification at high energies, where the structure of the resulting action may differ significantly. Second, the vanishing of the Hamiltonian indicates that these theories suffer from the problem of time, in which the Hamiltonian operator does not yield the time evolution of the quantum state of the field, a well-known problem in canonical approaches to quantum gravity \cite{Anderson:2010xm,Isham:1992ms}, though we note that in the framework of effective field theory, general relativity as a perturbative quantum field theory remains predictive at low energies \cite{Schwartz:2014sze}. In any case, the further development of the matter sector may be needed, particularly at the quantum level.

While we have shown that these approaches are in principle viable, the actions we have proposed are not austere in the sense that we have made no attempt to propose a minimal set of fields and symmetries (or other fundamental principles) to constrain the action in a unique manner, as is often done in the construction of field theories. Such constraints would be useful for limiting the allowed modifications to these theories, which in principle form a rather large class (in particular, one can modify the theories we have considered adding to Eqs. \eqref{Eq:TFEFE} and \eqref{Eq:CottonGravityCTensor} for instance any transverse and trace-free tensor $S_{ab}$ that one can construct from the metric and curvature). Of course, we again emphasize that the actions proposed here are not intended to be fundamental, and should be regarded as effective descriptions (formulated at the classical level). 

It may be that the approaches we have considered here ultimately trade austerity and completeness (as we do not expect our theories to be fundamental) for the avoidance of the extreme fine-tuning problem arising from vacuum energy contributions to the CC---some readers may regard this as too high a price to pay. On the other hand, there is some hope that perhaps one can reorganize the action and integrate out the appropriate degrees of freedom to obtain a more natural form for the action in which some principle or symmetry may be more easily identified. A fair question, which we have not addressed, concerns the mechanism that sets the value for the CC; while a fine-tuning in the early universe can be avoided so long as the field equations remain insensitive to the cosmological term, a mechanism for determining the initial data specifying the CC is needed. In short, our work shows that the class of approaches we have considered are viable according to some basic criteria, but at the same time, this approach illustrates some of the challenges and trade offs that one must make in order to avoid these problems. In any case, we hope that our work can provide some guidance for future exploration of scenarios in which the CC arises as an integration constant.


\section*{Acknowledgements}
We thank Marco de Cesare, Naresh Dadhich, and Victor Tapia for helpful comments and references. PC and JCF acknowledge support from the Taiwan National Science and Technology Council (NSTC) under project numbers
110-2112-M-002-031 and 112-2811-M-002-132, and also from the Leung Center for Cosmology and Particle Astrophysics (LeCosPA), National Taiwan University.


\bibliographystyle{elsarticle-harv} 
\bibliography{ref}

\begin{thebibliography}{59}
\expandafter\ifx\csname natexlab\endcsname\relax\def\natexlab#1{#1}\fi
\providecommand{\url}[1]{\texttt{#1}}
\providecommand{\href}[2]{#2}
\providecommand{\path}[1]{#1}
\providecommand{\DOIprefix}{doi:}
\providecommand{\ArXivprefix}{arXiv:}
\providecommand{\URLprefix}{URL: }
\providecommand{\Pubmedprefix}{pmid:}
\providecommand{\doi}[1]{\href{http://dx.doi.org/#1}{\path{#1}}}
\providecommand{\Pubmed}[1]{\href{pmid:#1}{\path{#1}}}
\providecommand{\bibinfo}[2]{#2}
\ifx\xfnm\relax \def\xfnm[#1]{\unskip,\space#1}\fi
\bibitem[{Alvarez and Velasco-Aja(2023)}]{Alvarez:2023utn}
\bibinfo{author}{Alvarez, E.}, \bibinfo{author}{Velasco-Aja, E.}, \bibinfo{year}{2023}.
\newblock \bibinfo{title}{{A Primer on Unimodular Gravity}}. \bibinfo{publisher}{Springer}.
\newblock pp. \bibinfo{pages}{1--43}.
\newblock \DOIprefix\doi{10.1007/978-981-19-3079-9_15-1}, \href{http://arxiv.org/abs/2301.07641}{{\tt arXiv:2301.07641}}.
\bibitem[{de~Alwis(2005)}]{deAlwis:2005tg}
\bibinfo{author}{de~Alwis, S.P.}, \bibinfo{year}{2005}.
\newblock \bibinfo{title}{{On integrating out heavy fields in SUSY theories}}.
\newblock \bibinfo{journal}{Phys. Lett. B} \bibinfo{volume}{628}, \bibinfo{pages}{183--187}.
\newblock \DOIprefix\doi{10.1016/j.physletb.2005.09.027}, \href{http://arxiv.org/abs/hep-th/0506267}{{\tt arXiv:hep-th/0506267}}.
\bibitem[{Anderson(2010)}]{Anderson:2010xm}
\bibinfo{author}{Anderson, E.}, \bibinfo{year}{2010}.
\newblock \bibinfo{title}{{The Problem of Time in Quantum Gravity}} \href{http://arxiv.org/abs/1009.2157}{{\tt arXiv:1009.2157}}. \bibinfo{note}{(preprint)}.
\bibitem[{Anderson and Finkelstein(1971)}]{Anderson:1971pn}
\bibinfo{author}{Anderson, J.L.}, \bibinfo{author}{Finkelstein, D.}, \bibinfo{year}{1971}.
\newblock \bibinfo{title}{{Cosmological constant and fundamental length}}.
\newblock \bibinfo{journal}{Am. J. Phys.} \bibinfo{volume}{39}, \bibinfo{pages}{901--904}.
\newblock \DOIprefix\doi{10.1119/1.1986321}.
\bibitem[{Barnes(2023)}]{Barnes:2023uru}
\bibinfo{author}{Barnes, A.}, \bibinfo{year}{2023}.
\newblock \bibinfo{title}{{Vacuum Static Spherically Symmetric Spacetimes in Harada's Theory}} \href{http://arxiv.org/abs/2309.05336}{{\tt arXiv:2309.05336}}.
\bibitem[{Barnes(2024)}]{Barnes:2024gko}
\bibinfo{author}{Barnes, A.}, \bibinfo{year}{2024}.
\newblock \bibinfo{title}{{Spherically symmetric electrovac spacetimes in conformal Killing gravity}}.
\newblock \bibinfo{journal}{Class. Quant. Grav.} \bibinfo{volume}{41}, \bibinfo{pages}{155007}.
\newblock \DOIprefix\doi{10.1088/1361-6382/ad5c33}.
\bibitem[{van Beest et~al.(2022)van Beest, Calder\'on-Infante, Mirfendereski and Valenzuela}]{vanBeest:2021lhn}
\bibinfo{author}{van Beest, M.}, \bibinfo{author}{Calder\'on-Infante, J.}, \bibinfo{author}{Mirfendereski, D.}, \bibinfo{author}{Valenzuela, I.}, \bibinfo{year}{2022}.
\newblock \bibinfo{title}{{Lectures on the Swampland Program in String Compactifications}}.
\newblock \bibinfo{journal}{Phys. Rept.} \bibinfo{volume}{989}, \bibinfo{pages}{1--50}.
\newblock \DOIprefix\doi{10.1016/j.physrep.2022.09.002}, \href{http://arxiv.org/abs/2102.01111}{{\tt arXiv:2102.01111}}.
\bibitem[{Beltran~Jimenez et~al.(2018)Beltran~Jimenez, Heisenberg, Olmo and Rubiera-Garcia}]{BeltranJimenez:2017doy}
\bibinfo{author}{Beltran~Jimenez, J.}, \bibinfo{author}{Heisenberg, L.}, \bibinfo{author}{Olmo, G.J.}, \bibinfo{author}{Rubiera-Garcia, D.}, \bibinfo{year}{2018}.
\newblock \bibinfo{title}{{Born\textendash{}Infeld inspired modifications of gravity}}.
\newblock \bibinfo{journal}{Phys. Rept.} \bibinfo{volume}{727}, \bibinfo{pages}{1--129}.
\newblock \DOIprefix\doi{10.1016/j.physrep.2017.11.001}, \href{http://arxiv.org/abs/1704.03351}{{\tt arXiv:1704.03351}}.
\bibitem[{van~der Bij et~al.(1982)van~der Bij, van Dam and Ng}]{vanderBij:1981ym}
\bibinfo{author}{van~der Bij, J.J.}, \bibinfo{author}{van Dam, H.}, \bibinfo{author}{Ng, Y.J.}, \bibinfo{year}{1982}.
\newblock \bibinfo{title}{{The Exchange of Massless Spin Two Particles}}.
\newblock \bibinfo{journal}{Physica A} \bibinfo{volume}{116}, \bibinfo{pages}{307--320}.
\newblock \DOIprefix\doi{10.1016/0378-4371(82)90247-3}.
\bibitem[{Buchmuller and Dragon(1988)}]{Buchmuller:1988wx}
\bibinfo{author}{Buchmuller, W.}, \bibinfo{author}{Dragon, N.}, \bibinfo{year}{1988}.
\newblock \bibinfo{title}{{Einstein Gravity From Restricted Coordinate Invariance}}.
\newblock \bibinfo{journal}{Phys. Lett. B} \bibinfo{volume}{207}, \bibinfo{pages}{292--294}.
\newblock \DOIprefix\doi{10.1016/0370-2693(88)90577-1}.
\bibitem[{Buchmuller and Dragon(1989)}]{Buchmuller:1988yn}
\bibinfo{author}{Buchmuller, W.}, \bibinfo{author}{Dragon, N.}, \bibinfo{year}{1989}.
\newblock \bibinfo{title}{{Gauge Fixing and the Cosmological Constant}}.
\newblock \bibinfo{journal}{Phys. Lett. B} \bibinfo{volume}{223}, \bibinfo{pages}{313--317}.
\newblock \DOIprefix\doi{10.1016/0370-2693(89)91608-0}.
\bibitem[{Carballo-Rubio et~al.(2022)Carballo-Rubio, Garay and Garc\'\i{}a-Moreno}]{Carballo-Rubio:2022ofy}
\bibinfo{author}{Carballo-Rubio, R.}, \bibinfo{author}{Garay, L.J.}, \bibinfo{author}{Garc\'\i{}a-Moreno, G.}, \bibinfo{year}{2022}.
\newblock \bibinfo{title}{{Unimodular gravity vs general relativity: a status report}}.
\newblock \bibinfo{journal}{Class. Quant. Grav.} \bibinfo{volume}{39}, \bibinfo{pages}{243001}.
\newblock \DOIprefix\doi{10.1088/1361-6382/aca386}, \href{http://arxiv.org/abs/2207.08499}{{\tt arXiv:2207.08499}}.
\bibitem[{de~Cesare and Wilson-Ewing(2022)}]{deCesare:2021wmk}
\bibinfo{author}{de~Cesare, M.}, \bibinfo{author}{Wilson-Ewing, E.}, \bibinfo{year}{2022}.
\newblock \bibinfo{title}{{Interacting dark sector from the trace-free Einstein equations: Cosmological perturbations with no instability}}.
\newblock \bibinfo{journal}{Phys. Rev. D} \bibinfo{volume}{106}, \bibinfo{pages}{023527}.
\newblock \DOIprefix\doi{10.1103/PhysRevD.106.023527}, \href{http://arxiv.org/abs/2112.12701}{{\tt arXiv:2112.12701}}.
\bibitem[{Chen(2010)}]{Chen:2010at}
\bibinfo{author}{Chen, P.}, \bibinfo{year}{2010}.
\newblock \bibinfo{title}{{Gauge Theory of Gravity with de Sitter Symmetry as a Solution to the Cosmological Constant Problem and the Dark Energy Puzzle}}.
\newblock \bibinfo{journal}{Mod. Phys. Lett. A} \bibinfo{volume}{25}, \bibinfo{pages}{2795--2803}.
\newblock \DOIprefix\doi{10.1142/S0217732310034274}, \href{http://arxiv.org/abs/1002.4275}{{\tt arXiv:1002.4275}}.
\bibitem[{Cl\'ement and Nouicer(2024a)}]{Clement:2023tyx}
\bibinfo{author}{Cl\'ement, G.}, \bibinfo{author}{Nouicer, K.}, \bibinfo{year}{2024}a.
\newblock \bibinfo{title}{{Cotton gravity is not predictive}}.
\newblock \bibinfo{journal}{Phys. Lett. B} \bibinfo{volume}{856}, \bibinfo{pages}{138947}.
\newblock \DOIprefix\doi{10.1016/j.physletb.2024.138947}, \href{http://arxiv.org/abs/2312.17662}{{\tt arXiv:2312.17662}}.
\bibitem[{Cl\'ement and Nouicer(2024b)}]{Clement:2024pjl}
\bibinfo{author}{Cl\'ement, G.}, \bibinfo{author}{Nouicer, K.}, \bibinfo{year}{2024}b.
\newblock \bibinfo{title}{{Farewell to Cotton gravity}} \href{http://arxiv.org/abs/2401.16008}{{\tt arXiv:2401.16008}}. \bibinfo{note}{(preprint)}.
\bibitem[{Cook(2008)}]{Cook:2008mx}
\bibinfo{author}{Cook, R.J.}, \bibinfo{year}{2008}.
\newblock \bibinfo{title}{{The Gravitational-Electromagnetic Analogy: A Possible Solution to the Vacuum-Energy and Dark-Energy Problems}} \href{http://arxiv.org/abs/0810.4495}{{\tt arXiv:0810.4495}}. \bibinfo{note}{(preprint)}.
\bibitem[{Dadhich(2017)}]{Dadhich:2016vbn}
\bibinfo{author}{Dadhich, N.}, \bibinfo{year}{2017}.
\newblock \bibinfo{title}{{Understanding General Relativity after 100 years: A matter of perspective}}.
\newblock \bibinfo{journal}{Fundam. Theor. Phys.} \bibinfo{volume}{187}, \bibinfo{pages}{73--87}.
\newblock \DOIprefix\doi{{10.1007/978-3-319-51700-1_7}}, \href{http://arxiv.org/abs/1609.02138}{{\tt arXiv:1609.02138}}.
\bibitem[{De~Felice and Tsujikawa(2010)}]{DeFelice:2010aj}
\bibinfo{author}{De~Felice, A.}, \bibinfo{author}{Tsujikawa, S.}, \bibinfo{year}{2010}.
\newblock \bibinfo{title}{{f(R) theories}}.
\newblock \bibinfo{journal}{Living Rev. Rel.} \bibinfo{volume}{13}, \bibinfo{pages}{3}.
\newblock \DOIprefix\doi{10.12942/lrr-2010-3}, \href{http://arxiv.org/abs/1002.4928}{{\tt arXiv:1002.4928}}.
\bibitem[{Deffayet et~al.(2022)Deffayet, Mukohyama and Vikman}]{Deffayet:2021nnt}
\bibinfo{author}{Deffayet, C.}, \bibinfo{author}{Mukohyama, S.}, \bibinfo{author}{Vikman, A.}, \bibinfo{year}{2022}.
\newblock \bibinfo{title}{{Ghosts without Runaway Instabilities}}.
\newblock \bibinfo{journal}{Phys. Rev. Lett.} \bibinfo{volume}{128}, \bibinfo{pages}{041301}.
\newblock \DOIprefix\doi{10.1103/PhysRevLett.128.041301}, \href{http://arxiv.org/abs/2108.06294}{{\tt arXiv:2108.06294}}.
\bibitem[{Duff and van Nieuwenhuizen(1980)}]{Duff:1980qv}
\bibinfo{author}{Duff, M.J.}, \bibinfo{author}{van Nieuwenhuizen, P.}, \bibinfo{year}{1980}.
\newblock \bibinfo{title}{{Quantum Inequivalence of Different Field Representations}}.
\newblock \bibinfo{journal}{Phys. Lett. B} \bibinfo{volume}{94}, \bibinfo{pages}{179--182}.
\newblock \DOIprefix\doi{10.1016/0370-2693(80)90852-7}.
\bibitem[{{Einstein}(1917)}]{Einstein:1917}
\bibinfo{author}{{Einstein}, A.}, \bibinfo{year}{1917}.
\newblock \bibinfo{title}{{Kosmologische Betrachtungen zur allgemeinen Relativit{\"a}tstheorie}}.
\newblock \bibinfo{journal}{Sitzungsberichte der K\&ouml;niglich Preussischen Akademie der Wissenschaften} , \bibinfo{pages}{142--152}.
\bibitem[{Einstein(1919)}]{Einstein:1919}
\bibinfo{author}{Einstein, A.}, \bibinfo{year}{1919}.
\newblock \bibinfo{title}{{Do gravitational fields play an essential part in the structure of the elementary particles of matter?}}
\newblock \bibinfo{journal}{Sitzungsber. Preuss. Akad. Wiss. Berlin (Math. Phys. )} \bibinfo{volume}{1919}, \bibinfo{pages}{349--356}.
\bibitem[{Ellis et~al.(2011)Ellis, van Elst, Murugan and Uzan}]{Ellis:2010uc}
\bibinfo{author}{Ellis, G.F.R.}, \bibinfo{author}{van Elst, H.}, \bibinfo{author}{Murugan, J.}, \bibinfo{author}{Uzan, J.P.}, \bibinfo{year}{2011}.
\newblock \bibinfo{title}{{On the Trace-Free Einstein Equations as a Viable Alternative to General Relativity}}.
\newblock \bibinfo{journal}{Class. Quant. Grav.} \bibinfo{volume}{28}, \bibinfo{pages}{225007}.
\newblock \DOIprefix\doi{10.1088/0264-9381/28/22/225007}, \href{http://arxiv.org/abs/1008.1196}{{\tt arXiv:1008.1196}}.
\bibitem[{Feng et~al.(2022)Feng, Mukohyama and Carloni}]{Feng:2022rga}
\bibinfo{author}{Feng, J.C.}, \bibinfo{author}{Mukohyama, S.}, \bibinfo{author}{Carloni, S.}, \bibinfo{year}{2022}.
\newblock \bibinfo{title}{{Junction conditions and sharp gradients in generalized coupling theories}}.
\newblock \bibinfo{journal}{Phys. Rev. D} \bibinfo{volume}{105}, \bibinfo{pages}{104036}.
\newblock \DOIprefix\doi{10.1103/PhysRevD.105.104036}, \href{http://arxiv.org/abs/2203.00011}{{\tt arXiv:2203.00011}}.
\bibitem[{Harada(2021)}]{Harada:2021bte}
\bibinfo{author}{Harada, J.}, \bibinfo{year}{2021}.
\newblock \bibinfo{title}{{Emergence of the Cotton tensor for describing gravity}}.
\newblock \bibinfo{journal}{Phys. Rev. D} \bibinfo{volume}{103}, \bibinfo{pages}{L121502}.
\newblock \DOIprefix\doi{10.1103/PhysRevD.103.L121502}, \href{http://arxiv.org/abs/2105.09304}{{\tt arXiv:2105.09304}}.
\bibitem[{Harada(2023)}]{Harada:2023rqw}
\bibinfo{author}{Harada, J.}, \bibinfo{year}{2023}.
\newblock \bibinfo{title}{{Gravity at cosmological distances: Explaining the accelerating expansion without dark energy}}.
\newblock \bibinfo{journal}{Phys. Rev. D} \bibinfo{volume}{108}, \bibinfo{pages}{044031}.
\newblock \DOIprefix\doi{10.1103/PhysRevD.108.044031}, \href{http://arxiv.org/abs/2308.02115}{{\tt arXiv:2308.02115}}.
\bibitem[{Henneaux and Teitelboim(1984)}]{Henneaux:1984ji}
\bibinfo{author}{Henneaux, M.}, \bibinfo{author}{Teitelboim, C.}, \bibinfo{year}{1984}.
\newblock \bibinfo{title}{{THE COSMOLOGICAL CONSTANT AS A CANONICAL VARIABLE}}.
\newblock \bibinfo{journal}{Phys. Lett. B} \bibinfo{volume}{143}, \bibinfo{pages}{415--420}.
\newblock \DOIprefix\doi{10.1016/0370-2693(84)91493-X}.
\bibitem[{Henneaux and Teitelboim(1989)}]{Henneaux:1989zc}
\bibinfo{author}{Henneaux, M.}, \bibinfo{author}{Teitelboim, C.}, \bibinfo{year}{1989}.
\newblock \bibinfo{title}{{The Cosmological Constant and General Covariance}}.
\newblock \bibinfo{journal}{Phys. Lett. B} \bibinfo{volume}{222}, \bibinfo{pages}{195--199}.
\newblock \DOIprefix\doi{10.1016/0370-2693(89)91251-3}.
\bibitem[{Isham(1993)}]{Isham:1992ms}
\bibinfo{author}{Isham, C.J.}, \bibinfo{year}{1993}.
\newblock \bibinfo{title}{{Canonical quantum gravity and the problem of time}}.
\newblock \bibinfo{journal}{NATO Sci. Ser. C} \bibinfo{volume}{409}, \bibinfo{pages}{157--287}.
\newblock \href{http://arxiv.org/abs/gr-qc/9210011}{{\tt arXiv:gr-qc/9210011}}.
\bibitem[{Kilmister(1962)}]{Kilmister:1959gqq}
\bibinfo{author}{Kilmister, C.W.}, \bibinfo{year}{1962}.
\newblock \bibinfo{title}{{The expression of field equations in terms of flux from sources}}.
\newblock \bibinfo{journal}{Colloq. Int. CNRS} \bibinfo{volume}{91}, \bibinfo{pages}{45--55}.
\bibitem[{Kilmister and Newman(1961)}]{Kilmister:1961}
\bibinfo{author}{Kilmister, C.W.}, \bibinfo{author}{Newman, D.J.}, \bibinfo{year}{1961}.
\newblock \bibinfo{title}{The use of algebraic structures in physics}.
\newblock \bibinfo{journal}{Mathematical Proceedings of the Cambridge Philosophical Society} \bibinfo{volume}{57}, \bibinfo{pages}{851–864}.
\newblock \DOIprefix\doi{10.1017/S0305004100036008}.
\bibitem[{Kim(2014)}]{Kim:2013nna}
\bibinfo{author}{Kim, H.C.}, \bibinfo{year}{2014}.
\newblock \bibinfo{title}{{Physics at the surface of a star in Eddington-inspired Born-Infeld gravity}}.
\newblock \bibinfo{journal}{Phys. Rev. D} \bibinfo{volume}{89}, \bibinfo{pages}{064001}.
\newblock \DOIprefix\doi{10.1103/PhysRevD.89.064001}, \href{http://arxiv.org/abs/1312.0705}{{\tt arXiv:1312.0705}}.
\bibitem[{Mantica and Molinari(2023a)}]{Mantica:2022flg}
\bibinfo{author}{Mantica, C.A.}, \bibinfo{author}{Molinari, L.G.}, \bibinfo{year}{2023}a.
\newblock \bibinfo{title}{{Codazzi tensors and their space-times and Cotton gravity}}.
\newblock \bibinfo{journal}{Gen. Rel. Grav.} \bibinfo{volume}{55}, \bibinfo{pages}{62}.
\newblock \DOIprefix\doi{10.1007/s10714-023-03106-7}, \href{http://arxiv.org/abs/2210.06173}{{\tt arXiv:2210.06173}}.
\bibitem[{Mantica and Molinari(2023b)}]{Mantica:2023stl}
\bibinfo{author}{Mantica, C.A.}, \bibinfo{author}{Molinari, L.G.}, \bibinfo{year}{2023}b.
\newblock \bibinfo{title}{{Note on Harada\textquoteright{}s conformal Killing gravity}}.
\newblock \bibinfo{journal}{Phys. Rev. D} \bibinfo{volume}{108}, \bibinfo{pages}{124029}.
\newblock \DOIprefix\doi{10.1103/PhysRevD.108.124029}, \href{http://arxiv.org/abs/2308.06803}{{\tt arXiv:2308.06803}}.
\bibitem[{Montesinos and Gonzalez(2023)}]{Montesinos:2023pjp}
\bibinfo{author}{Montesinos, M.}, \bibinfo{author}{Gonzalez, D.}, \bibinfo{year}{2023}.
\newblock \bibinfo{title}{{Diffeomorphism-invariant action principles for trace-free Einstein gravity}}.
\newblock \bibinfo{journal}{Phys. Rev. D} \bibinfo{volume}{108}, \bibinfo{pages}{124013}.
\newblock \DOIprefix\doi{10.1103/PhysRevD.108.124013}, \href{http://arxiv.org/abs/2312.03062}{{\tt arXiv:2312.03062}}.
\bibitem[{Montesinos and Gonzalez(2024)}]{Montesinos:2024yjs}
\bibinfo{author}{Montesinos, M.}, \bibinfo{author}{Gonzalez, D.}, \bibinfo{year}{2024}.
\newblock \bibinfo{title}{{Trace-free Einstein gravity as a constrained bigravity theory}}.
\newblock \bibinfo{journal}{Phys. Rev. D} \bibinfo{volume}{110}, \bibinfo{pages}{064057}.
\newblock \DOIprefix\doi{10.1103/PhysRevD.110.064057}.
\bibitem[{Nojiri and Odintsov(2011)}]{Nojiri:2010wj}
\bibinfo{author}{Nojiri, S.}, \bibinfo{author}{Odintsov, S.D.}, \bibinfo{year}{2011}.
\newblock \bibinfo{title}{{Unified cosmic history in modified gravity: from F(R) theory to Lorentz non-invariant models}}.
\newblock \bibinfo{journal}{Phys. Rept.} \bibinfo{volume}{505}, \bibinfo{pages}{59--144}.
\newblock \DOIprefix\doi{10.1016/j.physrep.2011.04.001}, \href{http://arxiv.org/abs/1011.0544}{{\tt arXiv:1011.0544}}.
\bibitem[{Nojiri et~al.(2017)Nojiri, Odintsov and Oikonomou}]{Nojiri:2017ncd}
\bibinfo{author}{Nojiri, S.}, \bibinfo{author}{Odintsov, S.D.}, \bibinfo{author}{Oikonomou, V.K.}, \bibinfo{year}{2017}.
\newblock \bibinfo{title}{{Modified Gravity Theories on a Nutshell: Inflation, Bounce and Late-time Evolution}}.
\newblock \bibinfo{journal}{Phys. Rept.} \bibinfo{volume}{692}, \bibinfo{pages}{1--104}.
\newblock \DOIprefix\doi{10.1016/j.physrep.2017.06.001}, \href{http://arxiv.org/abs/1705.11098}{{\tt arXiv:1705.11098}}.
\bibitem[{O'Hanlon(1972)}]{OHanlon:1972xqa}
\bibinfo{author}{O'Hanlon, J.}, \bibinfo{year}{1972}.
\newblock \bibinfo{title}{{Intermediate-range gravity - a generally covariant model}}.
\newblock \bibinfo{journal}{Phys. Rev. Lett.} \bibinfo{volume}{29}, \bibinfo{pages}{137--138}.
\newblock \DOIprefix\doi{10.1103/PhysRevLett.29.137}.
\bibitem[{Padmanabhan(2016)}]{Padmanabhan:2016eld}
\bibinfo{author}{Padmanabhan, T.}, \bibinfo{year}{2016}.
\newblock \bibinfo{title}{{The Atoms Of Space, Gravity and the Cosmological Constant}}.
\newblock \bibinfo{journal}{Int. J. Mod. Phys. D} \bibinfo{volume}{25}, \bibinfo{pages}{1630020}.
\newblock \DOIprefix\doi{10.1142/S0218271816300202}, \href{http://arxiv.org/abs/1603.08658}{{\tt arXiv:1603.08658}}.
\bibitem[{Padmanabhan and Padmanabhan(2014)}]{Padmanabhan:2014nca}
\bibinfo{author}{Padmanabhan, T.}, \bibinfo{author}{Padmanabhan, H.}, \bibinfo{year}{2014}.
\newblock \bibinfo{title}{{Cosmological Constant from the Emergent Gravity Perspective}}.
\newblock \bibinfo{journal}{Int. J. Mod. Phys. D} \bibinfo{volume}{23}, \bibinfo{pages}{1430011}.
\newblock \DOIprefix\doi{10.1142/S0218271814300110}, \href{http://arxiv.org/abs/1404.2284}{{\tt arXiv:1404.2284}}.
\bibitem[{Palti(2019)}]{Palti:2019pca}
\bibinfo{author}{Palti, E.}, \bibinfo{year}{2019}.
\newblock \bibinfo{title}{{The Swampland: Introduction and Review}}.
\newblock \bibinfo{journal}{Fortsch. Phys.} \bibinfo{volume}{67}, \bibinfo{pages}{1900037}.
\newblock \DOIprefix\doi{10.1002/prop.201900037}, \href{http://arxiv.org/abs/1903.06239}{{\tt arXiv:1903.06239}}.
\bibitem[{Pani and Sotiriou(2012)}]{Pani:2012qd}
\bibinfo{author}{Pani, P.}, \bibinfo{author}{Sotiriou, T.P.}, \bibinfo{year}{2012}.
\newblock \bibinfo{title}{{Surface singularities in Eddington-inspired Born-Infeld gravity}}.
\newblock \bibinfo{journal}{Phys. Rev. Lett.} \bibinfo{volume}{109}, \bibinfo{pages}{251102}.
\newblock \DOIprefix\doi{10.1103/PhysRevLett.109.251102}, \href{http://arxiv.org/abs/1209.2972}{{\tt arXiv:1209.2972}}.
\bibitem[{Pani et~al.(2013)Pani, Sotiriou and Vernieri}]{Pani:2013qfa}
\bibinfo{author}{Pani, P.}, \bibinfo{author}{Sotiriou, T.P.}, \bibinfo{author}{Vernieri, D.}, \bibinfo{year}{2013}.
\newblock \bibinfo{title}{{Gravity with Auxiliary Fields}}.
\newblock \bibinfo{journal}{Phys. Rev. D} \bibinfo{volume}{88}, \bibinfo{pages}{121502}.
\newblock \DOIprefix\doi{10.1103/PhysRevD.88.121502}, \href{http://arxiv.org/abs/1306.1835}{{\tt arXiv:1306.1835}}.
\bibitem[{Perlmutter et~al.(1999)}]{SupernovaCosmologyProject:1998vns}
\bibinfo{author}{Perlmutter, S.}, et~al. (\bibinfo{collaboration}{Supernova Cosmology Project}), \bibinfo{year}{1999}.
\newblock \bibinfo{title}{{Measurements of $\Omega$ and $\Lambda$ from 42 High Redshift Supernovae}}.
\newblock \bibinfo{journal}{Astrophys. J.} \bibinfo{volume}{517}, \bibinfo{pages}{565--586}.
\newblock \DOIprefix\doi{10.1086/307221}, \href{http://arxiv.org/abs/astro-ph/9812133}{{\tt arXiv:astro-ph/9812133}}.
\bibitem[{Riess et~al.(1998)}]{SupernovaSearchTeam:1998fmf}
\bibinfo{author}{Riess, A.G.}, et~al. (\bibinfo{collaboration}{Supernova Search Team}), \bibinfo{year}{1998}.
\newblock \bibinfo{title}{{Observational evidence from supernovae for an accelerating universe and a cosmological constant}}.
\newblock \bibinfo{journal}{Astron. J.} \bibinfo{volume}{116}, \bibinfo{pages}{1009--1038}.
\newblock \DOIprefix\doi{10.1086/300499}, \href{http://arxiv.org/abs/astro-ph/9805201}{{\tt arXiv:astro-ph/9805201}}.
\bibitem[{Rodrigues et~al.(2011)Rodrigues, de~O.~Salles, Shapiro and Starobinsky}]{Rodrigues:2011zi}
\bibinfo{author}{Rodrigues, D.C.}, \bibinfo{author}{de~O.~Salles, F.}, \bibinfo{author}{Shapiro, I.L.}, \bibinfo{author}{Starobinsky, A.A.}, \bibinfo{year}{2011}.
\newblock \bibinfo{title}{{Auxiliary fields representation for modified gravity models}}.
\newblock \bibinfo{journal}{Phys. Rev. D} \bibinfo{volume}{83}, \bibinfo{pages}{084028}.
\newblock \DOIprefix\doi{10.1103/PhysRevD.83.084028}, \href{http://arxiv.org/abs/1101.5028}{{\tt arXiv:1101.5028}}.
\bibitem[{Schwartz(2014)}]{Schwartz:2014sze}
\bibinfo{author}{Schwartz, M.D.}, \bibinfo{year}{2014}.
\newblock \bibinfo{title}{{Quantum Field Theory and the Standard Model}}.
\newblock \bibinfo{publisher}{Cambridge University Press}.
\bibitem[{Sotiriou(2009)}]{Sotiriou:2008ve}
\bibinfo{author}{Sotiriou, T.P.}, \bibinfo{year}{2009}.
\newblock \bibinfo{title}{{6+1 lessons from f(R) gravity}}.
\newblock \bibinfo{journal}{J. Phys. Conf. Ser.} \bibinfo{volume}{189}, \bibinfo{pages}{012039}.
\newblock \DOIprefix\doi{10.1088/1742-6596/189/1/012039}, \href{http://arxiv.org/abs/0810.5594}{{\tt arXiv:0810.5594}}.
\bibitem[{Sussman et~al.(2024a)Sussman, Mantica, Molinari and N\'ajera}]{Sussman:2024iwk}
\bibinfo{author}{Sussman, R.A.}, \bibinfo{author}{Mantica, C.A.}, \bibinfo{author}{Molinari, L.G.}, \bibinfo{author}{N\'ajera, S.}, \bibinfo{year}{2024}a.
\newblock \bibinfo{title}{{Response to a critique of ''Cotton Gravity''}} \href{http://arxiv.org/abs/2401.10479}{{\tt arXiv:2401.10479}}. \bibinfo{note}{(preprint)}.
\bibitem[{Sussman et~al.(2024b)Sussman, Mantica, Molinari and N\'ajera}]{Sussman:2024qsg}
\bibinfo{author}{Sussman, R.A.}, \bibinfo{author}{Mantica, C.A.}, \bibinfo{author}{Molinari, L.G.}, \bibinfo{author}{N\'ajera, S.}, \bibinfo{year}{2024}b.
\newblock \bibinfo{title}{{Second Response to the critique of ''Cotton Gravity''}} \href{http://arxiv.org/abs/2402.01992}{{\tt arXiv:2402.01992}}. \bibinfo{note}{(preprint)}.
\bibitem[{Sussman and Najera(2023)}]{Sussman:2023eep}
\bibinfo{author}{Sussman, R.A.}, \bibinfo{author}{Najera, S.}, \bibinfo{year}{2023}.
\newblock \bibinfo{title}{{Exact solutions of Cotton Gravity in its Codazzi formulation}} \href{http://arxiv.org/abs/2312.02115}{{\tt arXiv:2312.02115}}.
\bibitem[{Tapia and Ross(1998)}]{Tapia:1998qt}
\bibinfo{author}{Tapia, V.}, \bibinfo{author}{Ross, D.K.}, \bibinfo{year}{1998}.
\newblock \bibinfo{title}{{Conformal fourth-rank gravity, non-vanishing cosmological constant and anisotropy}}.
\newblock \bibinfo{journal}{Class. Quant. Grav.} \bibinfo{volume}{15}, \bibinfo{pages}{245--249}.
\newblock \DOIprefix\doi{10.1088/0264-9381/15/1/019}.
\bibitem[{Tapia et~al.(1996)Tapia, Ross, Marrakchi and Cataldo}]{Tapia:1996ue}
\bibinfo{author}{Tapia, V.}, \bibinfo{author}{Ross, D.K.}, \bibinfo{author}{Marrakchi, A.E.}, \bibinfo{author}{Cataldo, M.}, \bibinfo{year}{1996}.
\newblock \bibinfo{title}{{Renormalizable conformally invariant model for the gravitational field}}.
\newblock \bibinfo{journal}{Class. Quant. Grav.} \bibinfo{volume}{13}, \bibinfo{pages}{3261--3267}.
\newblock \DOIprefix\doi{10.1088/0264-9381/13/12/017}.
\bibitem[{Tapia and Ujevic(1998)}]{Tapia:1998as}
\bibinfo{author}{Tapia, V.}, \bibinfo{author}{Ujevic, M.}, \bibinfo{year}{1998}.
\newblock \bibinfo{title}{{Universal field equations for metric-affine theories of gravity}}.
\newblock \bibinfo{journal}{Class. Quant. Grav.} \bibinfo{volume}{15}, \bibinfo{pages}{3719--3729}.
\newblock \DOIprefix\doi{10.1088/0264-9381/15/11/028}, \href{http://arxiv.org/abs/gr-qc/0605132}{{\tt arXiv:gr-qc/0605132}}.
\bibitem[{Ventagli et~al.(2024)Ventagli, Pani and Sotiriou}]{Ventagli:2023moe}
\bibinfo{author}{Ventagli, G.}, \bibinfo{author}{Pani, P.}, \bibinfo{author}{Sotiriou, T.P.}, \bibinfo{year}{2024}.
\newblock \bibinfo{title}{{Incompatibility of gravity theories with auxiliary fields with the standard model}}.
\newblock \bibinfo{journal}{Phys. Rev. D} \bibinfo{volume}{109}, \bibinfo{pages}{044002}.
\newblock \DOIprefix\doi{10.1103/PhysRevD.109.044002}, \href{http://arxiv.org/abs/2306.12350}{{\tt arXiv:2306.12350}}.
\bibitem[{Weinberg(1989)}]{Weinberg:1988cp}
\bibinfo{author}{Weinberg, S.}, \bibinfo{year}{1989}.
\newblock \bibinfo{title}{{The Cosmological Constant Problem}}.
\newblock \bibinfo{journal}{Rev. Mod. Phys.} \bibinfo{volume}{61}, \bibinfo{pages}{1--23}.
\newblock \DOIprefix\doi{10.1103/RevModPhys.61.1}.
\bibitem[{Yang(1974)}]{Yang:1974kj}
\bibinfo{author}{Yang, C.N.}, \bibinfo{year}{1974}.
\newblock \bibinfo{title}{{Integral Formalism for Gauge Fields}}.
\newblock \bibinfo{journal}{Phys. Rev. Lett.} \bibinfo{volume}{33}, \bibinfo{pages}{445--447}.
\newblock \DOIprefix\doi{10.1103/PhysRevLett.33.445}.

\end{thebibliography}

\end{document}